\documentstyle[preprint,prl,aps]{revtex}
\begin{document}
\draft
\bigskip
\title
{Improved Theory of the Muonium Hyperfine Structure}
\author{T. Kinoshita and M. Nio}
\address{ Newman Laboratory of Nuclear Studies,
Cornell University, Ithaca, NY 14853 }
\maketitle
\begin{abstract}
Terms contributing to the hyperfine structure of the muonium ground
state at the level of few tenths of kHz have been evaluated.
The $\alpha^2 (Z\alpha)$ radiative correction has been calculated
numerically to the precision of 0.02 kHz.
Leading $\ln (Z\alpha )$ terms of order $\alpha^{4-n} (Z\alpha)^n ,
n=1,2,3,$ and some relativistic corrections have been evaluated
analytically.
The theoretical uncertainty is now reduced to 0.17 kHz.
At present, however, it is not possible to test QED to this
precision because of the 1.34 kHz uncertainty due to the muon mass.
\end{abstract}


\narrowtext
The hyperfine splitting of the muonium ground state is one of very
precisely measured quantities~\cite{mariam}:
\begin{equation}
  \Delta \nu (\mbox{exp})~ =~ 4~463~302.88~(16)~\mbox{kHz}~~~~~~
(0.036~\mbox{ppm}).        \label{meas}
\end{equation}
Currently new experiment is in progress to improve the measurement
of $\Delta \nu (\mbox{exp})$ and muon mass by a factor of five or
more \cite{hughes}.
This is very important for testing the validity of quantum
electrodynamics (QED) since $\Delta \nu$ can be calculated very
precisely in QED,
being relatively free from the effect of hadronic interaction.
The precision of such a test is limited at present by the uncertainty in
theoretical calculation, which may exceed 1 kHz.
This paper reports our result in which we have reduced this uncertainty
by nearly an order of magnitude.

As is well known, the bulk of the hyperfine splitting is given by the
Fermi formula
\begin{equation}
  E_F={16 \over 3}{(Z\alpha)}^2 c R_{\infty} {{m_e} \over {m_{\mu}}}
  \left [ 1+{{m_e} \over {m_\mu}} \right ]^{-3},     \label{EF}
\end{equation}
where $Z$ is the charge of the muon in units of the electron charge,
$R_{\infty}$ is the Rydberg constant for infinite nuclear mass,
and $m_e$ and $m_\mu$ are the electron and muon masses, respectively.
Of course $Z = 1$ for the muon, but it is kept
in the formula in order to distinguish the contribution of
binding effect (Z$\alpha$) from that of radiative correction ($\alpha$).

Many correction terms of both $\alpha$ and Z$\alpha$ type have been
calculated over 40 years.
It is customary to classify them
into three types: radiative non-recoil correction, pure recoil
correction, and radiative-recoil correction.
In addition there is a small weak interaction contribution.
Thus one may write
\begin{equation}
  \Delta \nu (\mbox{theory}) =\Delta \nu (\mbox{rad})
  + \Delta \nu (\mbox{recoil})
  + \Delta \nu (\mbox{rad-recoil})
  + \Delta \nu (\mbox{weak}) .    \label{theoryform}
\end{equation}
Conventionally, the effect of hadronic vacuum polarization is
included in $\Delta \nu (\mbox{rad-recoil})$.

Purely radiative terms of orders $\alpha (Z\alpha )$ and $\alpha
(Z \alpha)^2$ have been known for some time~\cite{SY}:
\begin{eqnarray}
  \Delta \nu (\mbox{rad})&=& (1 + a_{\mu} ) \left (  1 + {3 \over 2}
(Z \alpha )^2 + a_e + \alpha (Z \alpha )( \ln 2 - {5 \over 2} ) \right .
 \nonumber   \\
  & &\left . - {{8 \alpha (Z \alpha )^2} \over {3 \pi}} \ln (Z\alpha)
\left [ \ln (Z\alpha) - \ln 4 + {281 \over 480} \right ] \right .
  \nonumber  \\
  & &\left . + {{\alpha (Z\alpha )^2} \over {\pi}} (15.38 \pm 0.29 )
\right ) E_F.  \label{nonrecoil}
\end{eqnarray}
Here $a_e$ and $a_{\mu}$ are the anomalous magnetic moments of the
electron and muon, respectively.
The appearance of the factor $(1 + a_{\mu} )$ in (\ref{nonrecoil}) is
in accord with our definition of $E_F$ in (\ref{EF}).
The known recoil corrections add up to~\cite{SY}
\begin{eqnarray}
  \Delta \nu (\mbox{recoil}&)&= \left ( - {{3Z\alpha} \over {\pi}}
{{m_e m_{\mu}} \over {m_{\mu}^2 -m_e^2 }} \ln {m_{\mu} \over {m_e}}
\right . \nonumber \\
  & &\left .  + {{\gamma^2} \over {m_e m_{\mu}}} \left [ 2 \ln {{m_r}
\over {2\gamma}} - 6 \ln 2 + {65 \over 18} \right ] \right ) E_F ,
 \label{recoil}
\end{eqnarray}
where $\gamma \equiv Z \alpha m_r , m_r =  m_e m_{\mu} /(m_e + m_{\mu} )$.
The radiative-recoil contributions, which arise from both lepton lines and
vacuum polarizations, are given by
\begin{eqnarray}
  \Delta \nu (\mbox{rad-recoil}&)&= {{\alpha (Z\alpha)} \over \pi^2}
{m_e \over m_{\mu}} \left ( - 2 \ln^2 {m_{\mu} \over {m_e}} + {13
\over 12} \ln {m_{\mu} \over m_e} \right .  \nonumber  \\
  & &\left .  + {21 \over 2} \zeta (3) + \zeta (2) + {35 \over 9} +
(2.15 \pm 0.14 ) \right .  \nonumber   \\
  & &\left . + {\alpha \over \pi} \left [ - {4 \over 3} \ln^3 {{m_{\mu}}
\over m_e} + {4 \over 3} \ln^2 {{m_{\mu}} \over m_e}  + {\cal O} \left
( \ln {{m_{\mu}} \over m_e} \right ) \right ] \right ) E_F ,
    \label{radrecoil}
\end{eqnarray}
The $\alpha (Z\alpha )$ term is known exactly~\cite{SY,EKS0} except for
the hadronic vacuum polarization contribution (the $(2.15
\pm 0.14)$ term)~\cite{KF}.
The $\ln^3$ and $\ln^2$ parts of the $\alpha^2 (Z\alpha )$ term were
evaluated by Eides $et~al$. \cite{EKS1}.
Finally there is a small contribution due to the $Z^0$ exchange.
Our re-evaluation of the standard-model estimate~\cite{BF} gives
\begin{equation}
  \Delta \nu (\mbox{weak}) \simeq  0.065 ~\mbox{kHz}.   \label{weak}
\end{equation}

As is clear from these results
one must know the $\alpha^2 (Z\alpha)$ radiative correction
in order to improve the theoretical prediction further.
Fig. 1 shows typical diagrams contributing to this order.
Recently, terms represented by the diagrams (a) - (e) of Fig. 1 have
been evaluated by Eides $et~al$. \cite{EKS2}.
Their results are as follows:
\begin{eqnarray}
  \Delta \nu (\mbox{Fig.1(a)}) &=& {36 \over 35} {{\alpha^2 (Z\alpha )}
\over \pi} E_F  \nonumber  \\
&=& 0.567 ~\mbox{kHz} ,~~~~~~~~~~~~~~~~~~~~~~~~~~~~~~~~~~~~~~ \label{fig1a}
\end{eqnarray}
\begin{eqnarray}
  \Delta \nu (\mbox{Fig.1(b)}) &=& \left ( {224 \over 15}
\ln 2 - {38 \over 15}\pi -{118 \over 225} \right ) {{\alpha^2
(Z\alpha )} \over \pi} E_F  \nonumber   \\
&=& 1.030 ~\mbox{kHz} ,~~~~~~~~~~~~~~~~~~~~~~~~~~~~~~~~~~~~   \label{fig1b}
\end{eqnarray}
\begin{eqnarray}
  \Delta \nu (\mbox{Fig.1(c)})&=& \left (
 -{4 \over 3} z^2  -{{20\sqrt{5}} \over 9} z - {64 \over 45} \ln 2
 + {\pi^2 \over 9} + {1043 \over 675} + {3 \over 8} \right )
{{\alpha^2 (Z\alpha )} \over \pi} E_F  \nonumber   \\
&=& -0.369 ~\mbox{kHz} ,   \label{fig1c}
\end{eqnarray}
\begin{eqnarray}
\Delta \nu (\mbox{Fig.1(d)}) &=& -0.310~742 \cdots ~{{\alpha^2 (Z\alpha )}
 \over \pi} E_F  \nonumber   \\
&=& -0.171 ~\mbox{kHz},~~~~~~~~~~~~~~~~~~~~~~~~~~~~~~~~~~~~~~\label{fig1d}
\end{eqnarray}
where $z = \ln ((1 + \sqrt{5})/2)$.
The results (\ref{fig1a}), (\ref{fig1b}) and (\ref{fig1c}) are
analytic, while
(\ref{fig1d}) was evaluated numerically after reducing the integral
to one dimension.
We confirmed these results by an independent numerical calculation.
However, our purely numerical evaluation of Fig. 1(e):
\begin{eqnarray}
  \Delta \nu (\mbox{Fig.1(e)}) &=& -0.472~48~(9) {{\alpha^2
(Z\alpha )} \over \pi} E_F  \nonumber   \\
&=& -0.261 ~\mbox{kHz}  ~~~~~~~~~~~~~~~~~~~~~~~~~~~~~~~~~~~\label{ourfig1e}
\end{eqnarray}
disagrees with the semi-analytic result of Ref. \cite{EKS3}.
Recently Eides \cite{Eides} found an error in the Table after Eq. (23)
of Ref. \cite{EKS3}.
Their corrected value is in good agreement with (\ref{ourfig1e}).

Fig. 2 shows the complete set of Feynman diagrams of type (f) of Fig. 1,
which has not yet been evaluated.
The primary purpose of this paper is to report a preliminary result
of our calculation for all diagrams of Fig. 2 carried out in the
Feynman gauge:
\begin{eqnarray}
  \Delta \nu (\mbox{Fig.1(f)}) &=& (-0.63~\pm~0.04) {{\alpha^2
(Z\alpha )} \over \pi} E_F   \nonumber  \\
&=& -0.347~(0.022)~\mbox{kHz} , ~~~~~~~~~~~~~~~~~~~~~~~~~\label{newresult}
\end{eqnarray}
where the error is mainly due to the uncertainty in extrapolating
the integral to zero infrared cutoff.
Details of calculation will be reported elsewhere.
The complete $\alpha^2 (Z\alpha)$ correction is the sum of
(\ref{fig1a}) --- (\ref{fig1d}), (\ref{ourfig1e}), and (\ref{newresult}):
\begin{equation}
  \Delta \nu (\mbox{Fig.1}) ~=~ 0.449~(0.022)~\mbox{kHz} . ~~~\label{fig1}
\end{equation}

Recently we have received two preprints from Eides and his
collaborators \cite{EKS4}, which report the result of their calculation,
carried out in the Fried-Yennie gauge,
for part of the diagrams of Fig. 2.
We were able to compare our result for the sum of diagrams H17, H18
and H19 with theirs since this sum is invariant under the covariant
gauge transformation.
They are in perfect agreement although individually they have quite
different values due to different gauges.
To compare other diagrams, we have to wait for completion of their
calculation.

The remaining theoretical uncertainty in $\Delta \nu$ comes mainly from
terms of orders $\alpha^4$ and $\alpha^3 (m_e /m_{\mu})$.
Although these diagrams are of higher order than (\ref{fig1}), they may
have numerically comparable magnitudes due to the appearance
of $\ln (Z\alpha )$ and/or $\ln (m_{\mu} /m_e )$ factors.
Some of these contributions are known:\cite{Kar}
\begin{eqnarray}
  \delta \Delta \nu &=& -{8 \over 3} \left ({\alpha \over
{2\pi}}-2{m_e \over m_{\mu}} + {Z \over 4} {m_e \over m_{\mu}} \right )
   {\alpha \over \pi} (Z\alpha )^2 \ln^2 (Z\alpha ) E_F  \nonumber   \\
                &~&+ {17 \over 8} (Z\alpha )^4 E_F   \nonumber   \\
&=& 0.287 ~\mbox{kHz} , ~~~~~~~~~~~~~~~~~~~~~~~~~~~~~~~~~~~~\label{karsh}
\end{eqnarray}
where the first three terms come from the magnetic form factor
correction to the $\delta$-function potential $V_F$ whose expectation
value is $E_F$,
the reduced mass correction to the $\ln^2 (Z\alpha )$ terms of
(\ref{nonrecoil}),
and the ln $k$ part of the Salpeter term of the Lamb shift
\cite{Salpeter}, respectively.
The last term is a higher order Breit correction.

Here we report additional terms evaluated in the NRQED perturbation
theory \cite{CL}:
\begin{eqnarray}
  \Delta \nu (\alpha) &=& < V_{50} G V_F > + \cdots ,  \nonumber   \\
  \Delta \nu (\beta) &=& < V_{\rm{2-loop}} G V_F > + \cdots ,
 \nonumber   \\
  \Delta \nu (\gamma) &=& < V_S G V_F > + \cdots ,  \nonumber   \\
  \Delta \nu (\delta) &=& < V_{\rm{hfs}} G (K \mbox{or} D) > + \cdots  ,
 \nonumber   \\
  \Delta \nu (\epsilon) &=& a_e < V_{40} G V_F > + \cdots  ,
 \label{perturbation}
\end{eqnarray}
where $< \cdots >$ means the difference of the triplet and singlet
expectation values with respect to the non-relativistic wave function
of the muonium ground state.
$G$ is the Green's function of the non-relativistic electron in the
muon Coulomb potential, and $D$ and $K$ are the Darwin and
$k^4$-kinetic energy term.
$V_{50}$ and $V_{\rm{2-loop}}$ are $\delta$-function potentials
whose expectation values are the $A_{50}$ term of the Lamb-shift
energy with one virtual photon spanning over any number of Coulomb
photons \cite{SY2}
and the term arising from two spanning photons as well as the
vacuum-polarization effect \cite{SY3}, respectively.
$V_S$ is the $\delta$-function potential corresponding to the
part of the Salpeter term not included in (\ref{karsh}).
$V_{\rm{hfs}}$ is the effective hyperfine interaction potential
corresponding to the $\ln (m_{\mu} /m_e )$ term of (\ref{recoil}).
The last term of (\ref{perturbation}) comes from the $A_{40}$ part
of the Lamb shift \cite{SY2} and the electron anomaly correction
to $V_F$.

All these contributions have been evaluated analytically.
Their numerical values are
\begin{eqnarray}
  \Delta \nu (\alpha) &=& -0.381~\mbox{kHz} ,  \nonumber   \\
  \Delta \nu (\beta) &=& -0.007~\mbox{kHz} ,  \nonumber   \\
  \Delta \nu (\gamma) &=& -0.189~\mbox{kHz} ,  \nonumber   \\
  \Delta \nu (\delta) &=& -0.210~\mbox{kHz} ,  \nonumber   \\
  \Delta \nu (\epsilon) &=& ~~0.004~\mbox{kHz} .   \label{ptresults}
\end{eqnarray}
The term $\Delta \nu (\alpha)$ is of order $\alpha (Z\alpha)^3
\ln (Z\alpha ) $ and was obtained by Lepage \cite{GPL}.
The term $\Delta \nu (\beta)$ is of order $\alpha^2 (Z\alpha)^2
\ln (Z\alpha )$.
Unfortunately, this evaluation is incomplete since $V_{\rm{2-loop}}$ is
not yet fully known.
However, the contribution of remaining terms will not be much larger
than the above result.
The term $\Delta \nu (\gamma)$ is of order $(m_e / m_{\mu} )
(Z\alpha )^3 \ln (Z\alpha )$.
$\Delta \nu (\delta)$ and $\Delta \nu (\epsilon)$
are proportional to $(Z\alpha)^3 \ln (Z\alpha ) (m_e /m_{\mu} )
\ln (m_{\mu} / m_e )$ and $\alpha^2 (Z\alpha)^2 \ln (Z\alpha )$,
respectively.

The terms (\ref{karsh}) and (\ref{ptresults}) add up to
$-$ 0.496 kHz.
The uncertainty due to uncalculated terms will be about 0.05 kHz.
Including these estimates and using the value of $\alpha$,
$R_{\infty}$ and $m_{\mu} /m_e$ from Refs. \cite{cage}, \cite{nez} and
\cite{mariam}:
\begin{eqnarray}
  \alpha^{-1}&=&137.035~997~9~(32)~~~(0.024~\rm{ppm})    \nonumber   \\
  R_{\infty} &=&10 ~973~~731.568~30 ~(31)~ {\rm m}^{-1},  \nonumber   \\
  {{ m_{\mu} } \over { m_e }} &=& 206.768~259~(62),    \label{constants}
\end{eqnarray}
we find
\begin{equation}
  \Delta \nu (\mbox{theory}) ~=~ 4~463~302.63~(1.34)~(0.21)~(0.17)~
  \mbox{kHz}  ~,    \label{newtheory}
\end{equation}
where the first and second errors reflect the uncertainties in the
measurements of $m_{\mu}$ and $\alpha^{-1}$ listed in (\ref{constants}).
The third error is purely theoretical and dominated by the uncertainty
in the last $\alpha (Z\alpha)^2$ term of (\ref{nonrecoil}).
Further reduction of this error is very important and will be attempted
shortly.
The agreement between $\Delta \nu (\mbox{theory})$ and
$\Delta \nu (\mbox{exp})$ is excellent, the difference being
\begin{equation}
- 0.26~(0.16)~(1.34)~(0.21)~(0.17)~\mbox{kHz}  ,   \label{diff}
\end{equation}
where the first error is from the experiment (\ref{meas}) and the rest
are carried over from (\ref{newtheory}).

The result (\ref{newtheory}) is obtained using the value of $\alpha$
from (\ref{constants}) which is determined from the quantum Hall effect.
Actually, a more accurate value of $\alpha$ is known from the theory
and measurement of the electron anomalous magnetic moment, which
is~\cite{kinoshita}
\begin{equation}
  \alpha^{-1} = 137.035~992~22~(94) .    \label{electronamm}
\end{equation}
If one uses this instead of (\ref{constants}), one finds
\begin{equation}
  \Delta \nu (\mbox{theory})' ~=~ 4~463~303.00~(1.34)~(0.06)~(0.17)~
  \mbox{kHz}  ~.    \label{newtheory2}
\end{equation}

Alternately, if one assumes that the uncertainty due to $\alpha$ is
0.06 kHz and that QED is correct to 0.17 kHz, one can determine the
muon mass from (\ref{meas}) and (\ref{newtheory2}).
This leads to
\begin{equation}
  {{ m_{\mu} } \over { m_e }} = 206.768~275~(11) ,    \label{massratio}
\end{equation}
which is 5.6 times more accurate than the value quoted in
(\ref{constants}).
This precision is close to that expected from the new
direct measurement of $m_{\mu}$ \cite{hughes}.
With the new measurement of muonium hyperfine structure and further
improvement of theory, it will be possible to replace (\ref{massratio})
by an even better one.
Comparison of this result with the directly measured $m_{\mu}$ may be
regarded as an alternative way to test the validity of QED.

\acknowledgments

We thank G. P. Lepage, P. Labelle and late D. R. Yennie for useful
discussions.
Thanks are due to M. I. Eides for communicating their preliminary result
and pointing out some misprints and oversights in our paper.
We thank the hospitality of the National Laboratory for High Energy
Physics (KEK), Japan, where T. K. was visiting on sabbatical leave
from Cornell University and M. N. was a short term visitor.
This research is supported in part by the U. S. National Science
Foundation.
Part of numerical work was conducted at the Cornell National
Supercomputing Facility, which receives major funding from the US
National Science Foundation and the IBM Corporation, with additional
support from New York State and members of the Corporate Research
Institute.

\begin{figure}
\caption{Representative diagrams contributing to the $\alpha^2
(Z \alpha)$ radiative corrections to the muonium hyperfine structure
in which two virtual photons are exchanged between $e^-$ and $\mu^+$.
The muon is represented by $\times$.
\label{alldiagrams}}
\end{figure}
\begin{figure}
\caption{Two-photon exchange diagrams with fourth-order
radiative corrections on the electron line.
Diagrams which are related to these diagrams by time reversal are not
shown explicitly.
The muon is represented by $\times$.
\label{ourdiagrams}}
\end{figure}

\end{document}